\documentclass{article}
\usepackage{graphicx}
\usepackage{amsmath}
\usepackage{amsfonts}

\title{A Comparison of Fick and Maxwell-Stefan Diffusion Formulations in PEMFC Cathode Gas Diffusion Layers}
\author{Michael Lindstrom \thanks{Mathematics Department, University of British Columbia,
{\tt mlrtlm@math.ubc.ca}} \and Brian Wetton \thanks{Mathematics Department, University of British Columbia, {\tt wetton@math.ubc.ca}. Corresponding author. }}

\begin{document}

\maketitle

\begin{abstract}

This paper explores the mathematical formulations of Fick and Maxwell-Stefan diffusion in the context of polymer electrolyte membrane fuel cell cathode gas diffusion layers. Formulations of diffusion combined with mass-averaged Darcy flow are considered for three component gases. 
Fick formulations can be considered as approximations of Maxwell-Stefan in a certain sense. 
For this application, the formulations can be compared computationally in a simple, one dimensional setting.  We observe that the predictions of the formulations are very similar, despite their seemingly different structure. Analytic insight is given to the result. In addition, it is seen that for both formulations, diffusion laws are small perturbations from bulk flow. The work is also intended 
as a reference to multi-component gas diffusion formulations in the fuel cell setting. 
\end{abstract}

\section{Introduction}

Polymer Electrolyte Membrane Fuel Cells (PEMFC) are promising energy producing electrochemical devices \cite{wilkinson}. They are very efficient and are non-polluting, having only water as a by-product, when pure Hydrogen is used as fuel. Reactant gases (oxidant on the cathode and hydrogen on the anode) are pumped through these devices, often in small channels. From these channels, gases are transported through Gas Diffusion Layers (GDL) to reactant sites in catalyst layers. GDL can be made of teflonated carbon fibre paper. 
Because these layers also often transport liquid water, they have also been called Porous Transport Layers (PTL) \cite{ptl}. Thus, models of PEMFC include a description of multi-component gas flow in porous media. There are many computational models describing aspects of fuel cell operation. We cite a two recent such models: \cite{fick} includes a Fick model of  multi-component gas diffusion and \cite{ms} a Maxwell-Stefan diffusion model. It is well known that (diagonal) Fick diffusion is an approximation of Maxwell-Stefan diffusion and only coincides with it when the molar masses and the binary diffusivities of the gas components  are identical \cite{stockiepaper}. However, in this research area it is known that the use of the Fick formulation can lead to small errors in modelling transport in the cathode GDL \cite{fms}. In this work, we compare the formulations in a simple, analytic framework that is representative of dry cathode GDL transport. We show that the two formulations do indeed give quite similar results, varying by only a few percent. We give analytic insight into why the results are so similar even though the structure of the formulations are so different. 

Complete modelling of cathode GDL transport should include liquid water. We consider dry transport to be able to focus on the question of ``Fick versus Stefan Maxwell" formulations, the purpose of this work. The two formulations should give quite different results on the anode, where Hydrogen has a very different molar mass and binary diffusivity, and is the majority phase if pure Hydrogen is used. However, in this case, Hydrogen transport in dry gas is so efficient it is rarely a limiting factor in PEMFC operation, which is why the anode is not considered in this study. Hydrogen starvation due to blockage by liquid water is quite significant, however, and can be a source of catalyst layer degradation \cite{review1}. 

In section~\ref{s:formulations} below, we present a summary of formulations of (consistent) Fick and Maxwell-Stefan diffusion. This is followed in section~\ref{s:application} by an application to PEMFC Cathode GDL conditions, where it is seen that a diagonal Fick approximation is appropriate. This is confirmed in a one-dimensional computational setting in section~\ref{s:computing}. All parameters used in the study are summarized in table~\ref{tab:constants}.

\section{Formulations}
\label{s:formulations} 

\subsection{Choice of unknowns}
\label{s:AB}

In this paper, we will model the concentration profiles within the gas diffusion layer of the cathode with three gas species: O$_2$, H$_2$O$_{\text{vap}}$ (water vapor), and N$_2$. Our model will be an isothermal (constant temperature), steady-state approximation to the concentration profiles of the gases. It is known that the time scale for gas dynamics in GDLs is extremely fast \cite{keith} so the steady state assumption is valid. The isothermal assumption is made just to simplify the presentation. We will let  ${\bf C} = (C_1, C_2, C_3)^T$ be the vector representing the molar concentration of the three gas species, where $C_i$ represents the concentration of species $i$ ($i=1$ for Oxygen, $i=2$ for water vapour, $i=3$ for Nitrogen). We will also denote the total concentration of all three gas species by $||{\bf C}|| = C_1 + C_2 + C_3.$ The concentrations have a bulk (mass-averaged) velocity $U$ and have molar diffusive fluxes ${\bf J}$ (in general nine scalar quantities, for each gas component in each coordinate direction) relative to that velocity. 
For this application $U$ will be determined by Darcy's Law for porous media flow, described below. Conservation of each component is expressed in the following three equations:  
\begin{equation}
\label{eq:ccons} 
\nabla \cdot (C_i U + J_i) = 0.
\end{equation}
Equivalently, we would define $Q_i = C_i U +J_i$, the total flux and write 
\begin{equation}
\label{eq:qcons} 
\nabla \cdot Q_i = 0.
\end{equation}
We can also write the conservation of total mass 
\begin{equation}
\label{eq:totcons}
\nabla \cdot (\rho U) = 0  
\end{equation}
where $\rho$ is the mass density given by $\rho = M_1 C_1 + M_2 C_2 + M_3 C_3$ where $M_i$ is the molar mass of species $i$. For (\ref{eq:ccons}) to be consistent with (\ref{eq:totcons}) we must have 
\begin{equation}
\label{eq:Jcons} 
M_1 J_1 + M_2 J_2 + M_3 J_3 = 0 
\end{equation}
in each direction and for (\ref{eq:qcons}) to be consistent with (\ref{eq:totcons}) 
\begin{equation}
\label{eq:Qcons} 
M_1 Q_1 + M_2 Q_2 + M_3 Q_3 = \rho U. 
\end{equation}
There are two formulations that can be used to describe the combined bulk and diffusive transport of the three species. They are equivalent as long as the consistency conditions above are met. 
\begin{description}
\item[A:] The three components of $\bf C$ are taken as unknowns with (\ref{eq:ccons}) or (\ref{eq:qcons}) as the equations. The resulting total density satisfies (\ref{eq:totcons}) automatically. 
\item[B:] Some choice of two components of $\bf C$ (say $C_1$ and $C_2$) and $\rho$ are taken as unknowns satisfying the corresponding two equations of (\ref{eq:ccons}) and (\ref{eq:totcons}). The remaining third component
\begin{equation}
\label{eq:c3}
C_3 = (\rho - M_1C_1 - M_2C_2)/M_3
\end{equation}
will automatically satisfy the remaining third equation of (\ref{eq:ccons}). 
\end{description}
Typically, computer implementations have been made using the formulation {\bf B} above since the velocity $U$ is used in total mass conservation (\ref{eq:totcons}) in many commercial and freeware codes for fluid flow. 

\subsection{Diffusion} 
\label{s:diff}

Maxwell-Stefan diffusion fluxes are determined by \cite{TK} (page 19) 
\begin{equation}
\label{eq:AJX}
\sum_j \mathcal{A}_{ij} J_j = \nabla \left( \frac{C_i}{|{\bf C}|} \right)
\end{equation}
where 
\[
\mathcal{A}_{ij} ({\bf C}) = \frac{1}{||{\bf C}||^2} \begin{cases} \sum_{\ell \neq i} \frac{-C_\ell}{D_{i \ell}} \text{ if } i=j \\ \frac{C_i}{D_{ij}} \text{ if } i \neq j \end{cases}.
\]
Note that (\ref{eq:AJX}) is a tensor equation, a $3 \times 3$ system with matrix $\mathcal{A}$ for the fluxes in each coordinate direction. The binary diffusivities $D_{i,j}$ are given constants. 

Consider now (\ref{eq:AJX}) in a single coordinate direction $x$. The right hand side vector $\bf G$ is given by 
\[
G_i = \frac{\partial}{\partial x} \left(\frac{C_i}{||{\bf C}||}\right)
\]
and the system for the $x$ component of the fluxes $\bf J$ is (\ref{eq:AJX}) written in matrix form:
\begin{equation}
\label{eq:AJX2}
\mathcal{A} {\bf J} = {\bf G}. 
\end{equation}
The matrix $\mathcal A$ is {\em not invertible} (it has rank 2). The system (\ref{eq:AJX2}) is solvable only when the right hand satisfies 
\begin{equation}
\label{eq:project}
G_1 + G_2 + G_3 = 0 
\end{equation}
but that is always true for the form of vector $\bf X$ taken as the right hand side. In mathematical terms, the condition above is the standard one for rank deficient systems, that $\bf X$ must be perpendicular to the nullspace of $\mathcal{A}^T$, which is spanned by $[1, 1, 1]^T$. The solution fluxes $\bf J$ are not determined uniquely. They are determined up to a multiple 
of the nullvector  $C = [C_1, C_2, C_3]^T$ of $\mathcal{A}$. The arbitrary multiple has the physical significance of a molar averaged velocity, which should be chosen so that the mass average of the fluxes (\ref{eq:Jcons}) is satisfied. 

There are a number of ways to proceed to write modified linear systems for $\bf J$ of full rank. The first approach, suitable for formulation {\bf A} of section~\ref{s:AB} is to augment $\bf J$ with a scalar $\xi$ and solve the following $4 \times 4$ full rank system 
\begin{equation}
\label{eq:ahatj}
\hat{\mathcal{A}}
\left[ \begin{array}{c} J_1 \\ J_2 \\ J_3 \\ \xi \end{array} \right] = 
\left[ \begin{array}{c} G_1 \\ G_2 \\ G_3 \\ 0 \end{array} \right] . 
\end{equation}
where (written in block form): 
\[
\hat{\mathcal{A}} = \left[
\begin{array}{ccc|c}
 & & & 1 \\
& \mathcal{A} & & 1 \\
& & & 1 \\ \hline
M_1 & M_2 & M_3 & 0 
\end{array}
\right]. 
\]
The variable $\xi$ has the interpretation of a projection distance of the first three components of the right hand side onto the subspace (\ref{eq:project}). As long as the right hand is consistent, the resulting $\xi$ will be zero. This formulation could be useful in numerical approximations in which  (\ref{eq:project}) is only approximately satisfied. The same approach can be used to compute the total fluxes $\bf Q$:
\begin{equation}
\label{eq:ahatq}
\hat{\mathcal{A}}
\left[ \begin{array}{c} Q_1 \\ Q_2 \\ Q_3 \\ \xi \end{array} \right] = 
\left[ \begin{array}{c} G_1 \\ G_2 \\ G_3 \\ \rho U \end{array} \right] . 
\end{equation}
In principle, one could write an analytic expression for $\hat{\mathcal{A}}^{-1}$ but it is more practical to solve (\ref{eq:ahatj}) or (\ref{eq:ahatq}) numerically. 

We proceed to a second approach that is appropriate for formulation {\bf B} of  section~\ref{s:AB}, in which the variable $C_3$ is eliminated in favour of the total density $\rho$. Considering again (\ref{eq:AJX2}), it is clear that the first two equations contain all the information of this rank 2 system. They can be augmented with the mass averaged velocity condition (\ref{eq:Jcons}) to make a full rank $3 \times 3$ system: 
\begin{equation}
\label{eq:bfick}
\mathcal{B}
\left[ \begin{array}{c} J_1 \\ J_2 \\ J_3 \end{array} \right] = 
\left[ \begin{array}{c} G_1 \\ G_2 \\  0 \end{array} \right] . 
\end{equation}
where 
\[
\mathcal{B} = \frac{1}{||{\bf C}||^2} \left[
\begin{array}{ccc}
 -\frac{C_2}{D_{12}} -\frac{C_3}{D_{13}} & \frac{C_1}{D_{12}} & \frac{C_1}{D_{13}} \\
\frac{C_2}{D_{12}} &  -\frac{C_1}{D_{12}} -\frac{C_3}{D_{23}} &  \frac{C_2}{D_{23}}  \\
M_1 & M_2 & M_3
\end{array}
\right]. 
\]
where to complete the formulation, the terms $C_3$ above are replaced using (\ref{eq:c3}). 
This is a summary of a process that is described in two stages in \cite{stockiepaper}. 
In this formulation, only $J_1$ and $J_2$ are required so only the upper $2 \times 2$ block of $\mathcal{B}^{-1}$ is required. We call this block $- \mathcal{F}$ (it depends on $C_1$, $C_2$ and $\rho$) and write 
\begin{equation}
\label{eq:ffick}
\left[ \begin{array}{c} J_1 \\ J_2 \end{array} \right] = - \mathcal{F} 
\left[ \begin{array}{c} G_1 \\ G_2 \end{array} \right]. 
\end{equation}
This could be called a Fick formulation in the sense that $C_3$ has been eliminated. In fact, the similar procedure for fluxes relative to a molar averaged velocity leads to analytic expressions in \cite{TK} (page 80) that are called Fick fluxes. However, it should be noted that the fluxes $\bf J$ from (\ref{eq:ffick}) or equivalently from (\ref{eq:bfick}) are exactly the same as the Maxwell-Stefan fluxes computed from the system (\ref{eq:ahatj}). What is typically called Fick diffusion involves the diagonal approximation of the matrix $\mathcal{F}$ which is considered in a representative fuel cell setting below. 

The entries of $\mathcal{F}$ can be computed analytically: 
\begin{eqnarray}
\label{eq:f11}
\mathcal{F}_{11} &  = & ||{\bf C}||^2 \frac{(M_3 C_1 D_{23}+M_3 C_3 D_{12}+M_2 C_2 D_{12}) D_{13}}{
\rho (C_1 D_{23}+ C_2 D_{13}+ C_3 D_{12} )} \\
\label{eq:f12}
\mathcal{F}_{12} &  = & -||{\bf C}||^2 \frac{(M_2D_{12} - M_3D_{13}) C_1 D_{23}}{
\rho (C_1 D_{23}+ C_2 D_{13}+ C_3 D_{12} )} \\
\label{eq:f21}
\mathcal{F}_{21} &  = & -||{\bf C}||^2 \frac{(M_1D_{12} - M_3D_{23}) C_2 D_{13}}{
\rho (C_1 D_{23}+ C_2 D_{13}+ C_3 D_{12} )} \\
\label{eq:f22} 
\mathcal{F}_{22} &  = & +||{\bf C}||^2 \frac{(M_1 C1 D_{12}+M_3 C_2 D_{13}+M_3 C_3 D_{12}) D_{23}}{
\rho (C_1 D_{23}+ C_2 D_{13}+ C_3 D_{12} )}.
\end{eqnarray}
Above, $\rho$ has been introduced explicitly in some terms but $C_3$ left instead of replacing it by (\ref{eq:c3}) to simplify the expressions.Ê

\section{Application to PEMC GDL Cathodes}
\label{s:application}

\subsection{Cathode Conditions of a PEMFC}
\label{s:diagonal}

We fix the temperature at $350$ K at a pressure of $2$ barg, standard conditions for an 
older Ballard Mk 9 stack with which the authors are familiar \cite{stack}. The conclusions of the later analysis are not dependent on this exact choice of operating parameters. We assume the ideal gas law for the cathode gas mixture 
\begin{equation} 
 \label{ideal} 
 P = ||C|| RT 
\end{equation}
with ideal gas constant $R$. This yields the total molar concentration of gases to be $104$ mol m$^{-3}.$ At the prescribed temperature, we find the saturation pressure of water to be $4.17 \times 10^4$ Pa \cite{Steam},  which gives a water vapour pressure of $3.13 \times 10^{4}$ Pa at 75 \% humidity. This corresponds to a molar concentration of $10.7$ mol/m$^3.$ We assume the fuel cell is using ambient air, where within dry air the Oxygen concentration is approximately $21 \%$ and the Nitrogen concentration is approximately $79 \%.$ This yields cathode inlet concentrations of Oxygen and Nitrogen of $19.7$ and $74.0$ mol/m$^3$ respectively. The binary diffusivities $D_{ij}$ are temperature and pressure dependent \cite{TK}. Based on \cite{CRC}, we linearly interpolated the relations based on experimental data and used the approximate relation that the diffusivities are inversely proportional to pressure to obtain the values of the diffusivities in our work. A table summarizing all the constants and parameters of our study is given in table \ref{tab:constants}. 

Using the values in the table, we can compute the entries of the Fick matrix $\mathcal{F}$ (\ref{eq:f11}-\ref{eq:f22}):
\[
\mathcal{F} \approx 
\left[
\begin{array}{cc} 
1.20 \times 10^{-3} &   6.79 \times 10^{-5} \\
    -4.89 \times 10^{-5} &   1.04 \times 10^{-3} 
\end{array}
\right]
\]
To a good approximation, $\mathcal{F}$  can be taken as a multiple of the identity matrix. This fits in the framework of ``true" Fick diffusion where $\mathcal{F}$ is approximated by 
\[
||{\bf C}|| D
\]
times the identity matrix. Here, we have identified 
\[
D = (1.20+1.04)\times 10^{-3}/(2 ||{\bf C}||) \approx  
1.07 \times 10^{-5} \mbox{\ m$^2$ s$^{-1}$}
\]
in this model as a very good fit to true Maxwell-Stefan diffusion in this scenario. This is reinforced by the model computational results below. Several things should be noted at this point. The resulting fluxes $\bf J$ using the process (\ref{eq:ffick}) and the resulting consistent density equation (\ref{eq:totcons}) do not depend on which concentration ($C_3$, Nitrogen, in the case above) is replaced by the density. However, in the diagonal approximation above, there is a significant difference depending on which concentration is removed. If $C_2$ (vapour) is removed for example, the resulting $D \approx 8.99 \times 10^{-6}$ m$^2$ s$^{-1}$ and the off diagonal terms are significantly larger, up to half this value. The Fick model 
\[
{\bf J} = -||{\bf C}|| D \nabla \left( \frac{C_i}{|{\bf C}|} \right)
\]
with a single value of $D$ is only fully consistent with Maxwell-Stefan when all molar masses are identical (or equivalently if the fluxes are relative to a given {\em molar} averaged velocity) {\em and} all the binary diffusivities $D_{ij}$ are identical and equal to $D$. Considering the numerators of
(\ref{eq:f12}-\ref{eq:f21}) in the case of $D_{ij}$ and $M_i$ being of roughly comparable size as in our case, the off diagonal terms are reduced by taking $C_1$ and $C_2$ the minority phases ($C_3$ the largest concentration). This is done in our ordering and is the reason why the diagonal approximation in this case is so accurate. In addition, keeping $C_1$ (Oxygen concentration) and  $C_2$ (vapour concentration) as unknowns in the fuel cell setting makes sense since these quantities affect performance while Nitrogen is inert. 

\begin{table}
\centering
\begin{tabular}{l | c | r}
Density  & Value \\
\hline
Temperature & $T$ & $350$ K \\
Fuel cell pressure & $P$ & $3.04 \times 10^5$ Pa \\
Humidity & $H$ & $0.75$ \\
Current density & $I$ & $10^4$ A m$^{-2}$ \\
Porosity & $\phi $ & $0.74$ \\
Permeability & $\kappa$ & $10^{-15}$ m$^2$ \\
Viscosity & $\mu$ & $2.24 \times 10^{-5}$ kg m$^{-1}$ s$^{-1}$ \\
Binary diffusivity Oxygen-Water & $D_{12}$ & $1.19 \times 10^{-5}$ m$^2$ s$^{-1}$ \\
Binary diffuisivity Oxygen-Nitrogen & $D_{13}$ & $1.18 \times 10^{-5}$ m$^2$ s$^{-1}$ \\
Binary diffusivity Water-Nitrogen & $D_{23}$ & $9.23 \times 10^{-6}$ m$^2$ s$^{-1}$ \\
Fick diffusivity & $D$ & $1.07 \times 10^{-5}$ m$^2$ s$^{-1}$ \\
Length & $L$ & $2.5 \times 10^{-4}$ m \\
Oxygen concentration & $C_1$ & $19.7$ mol m $^{-3}$ \\
Water vapor concentration & $C_2$ & $10.7$ mol m $^{-3}$ \\
Nitrogen concentration & $C_3$ & $74.0$ mol m $^{-3}$ \\
Faraday's Constant & $F$ & $9.649 \times 10^4$ C mol$^{-1}$ \\
Oxygen molar mass & $M_1$ & $32\times 10^{-3}$ kg mol$^{-1}$ \\
Water molar mass & $M_2$ & $18\times 10^{-3}$ kg mol$^{-1}$ \\
Nitrogen molar mass & $M_3$ & $28\times 10^{-3}$ kg mol$^{-1}$ \\
Oxygen flux & $Q_1$ & $2.59 \times 10^{-2}$ mol m$^{-2}$ s$^{-1}$ \\
Water flux & $Q_2$ & $-5.18 \times 10^{-2}$ mol m$^{-2}$ s$^{-1}$ \\
Nitrogen flux & $Q_3$ & $0$ mol m$^{-2}$ s$^{-1}$ \\
Ideal gas constant & $R$ & $8.314$ kg m$^2$ mol$^{-1}$ s$^{-2}$ K$^{-1}$ \\
\end{tabular}
 \caption{Constants and parameters for dry PEMFC cathode transport.}
\label{tab:constants}
\end{table}

\subsection{Darcy's Law and Operating Fluxes}
\label{s:darcyop}

Darcy's law for porous media states that 
\[
U = -\frac{\kappa}{\phi \mu} \nabla P 
\]
for a permeability $\kappa,$ porosity $\phi$ and viscosity $\mu$. This can be combined with the ideal gas law (\ref{ideal}) to yield: 
\begin{equation} 
\label{eq:darcy}
U = -\frac{\kappa RT}{\phi \mu} \nabla ||C||. 
\end{equation} 
It is accepted that this velocity is a mass averaged velocity \cite{whittaker}. 
We absorb the coefficients of the gradient above into a single constant 
\[
\sigma = \frac{\kappa RT}{\phi \mu}.
\]
Variations of $\mu$ with concentration (and temperature) could be considered in a full model, but since we are interested in formulations of diffusion in this work, we will leave it as a representative constant. 

A typical current density drawn from the fuel cell is $I = 10^4$ A/m$^2$ (1 A/cm$^2$). The cathode reaction is 
\[
O_2 + 4p^+ + 4e^- \rightarrow 2H_2O.
\]
Thus the specified current $I$ corresponds to an oxygen flux of 
\[
Q_1 =  \frac{I}{4 F} \approx 0.0259 \mbox{\ mol/s/m$^2$} 
\]
through the GDL from channel to catalyst sites, where $F$ is Faraday's constant. 
In the scenario below we consider we consider that all product water returns to the cathode channels
\[
Q_2 \approx  -0.0518 \mbox{\ mol/s/m$^2$}.Ê
\]
As Nitrogen does not react, $Q_3=0$.

\section{Computing Cathode GDL Gradients from Fluxes}
\label{s:computing} 

We consider now a one dimensional profile of concentrations $C_1(x)$, $C_2(x)$ and $C_3(x)$ through the cathode GDL from $x=0$ (gas channel boundary) to $x=L$ (catalyst layer boundary). We will compute the vector ${\bf C}^\prime(0)$,  which we will denote $\bf c$, below using the formulas derived above. In other work \cite{keith,frank2} it is shown with a scaling argument that ${\bf C}^\prime(x)$ is constant on $[0,L]$ to leading order. This result applies to this case as well, so the resulting $\bf c$ describes accurately the concentration derivative across the GDL. 

We begin with the Maxwell-Stefan formulation (\ref{eq:ahatq}). We can write 
\[
G_1 = \frac{d}{dx} \frac{C_1}{||{\bf C}||} (0) = \frac{1}{C^2} (C c_1-C_1(c_1 + c_2 + c_3))
\]
where we have used $C$ as shorthand for $||{\bf C}|| = C_1 + C_2 + C_3$. Similar expressions for $C_2$ and $C_3$ can be combined into the expression
\[
{\bf G} = \mathcal{M} {\bf c}  
\]
where $\mathcal{M}$ is the $3 \times 3$ matrix 
\[
\frac{1}{C^2}
\left[
\begin{array}{ccc} 
C-C_1 &   -C_1  & -C_1 \\
-C_2 & C-C_2 & -C_2 \\
-C_3  & - C_3 & C-C_3 
\end{array}
\right]
\]
From Darcy's Law (\ref{eq:darcy}) we have 
\[
U = -\sigma (c_1 + c_2 + c_3) 
\]
Combining this in (\ref{eq:ahatq}) leads to the following system 
\[
\left[
\begin{array}{ccc|c}
 & & & -1 \\
& \mathcal{M} & & -1 \\
& & & -1 \\ \hline
-\rho \sigma & -\rho \sigma & -\rho \sigma & 0
\end{array}
\right]
\left[ \begin{array}{c} c_1 \\ c_2 \\ c_3 \\ \xi \end{array} \right] = 
\left[
\begin{array}{ccc}
 & & \\
& \mathcal{A} &  \\
& & \\ \hline
M_1 & M_2 & M_3
\end{array}
\right]
\left[ \begin{array}{c} Q_1 \\ Q_2 \\ Q_3 \end{array} \right]
\]
where $\xi$ has the same role as in (\ref{eq:ahatj}). 
The equation above is (\ref{eq:ahatq}) with the right and left hand sides exchanged and the $\xi$ terms moved to the other side of the equation. The matrix on the right hand side has size $4 \times 3$. 
The results with the values in Table~\ref{tab:constants} are given below:
\begin{equation}
\label{eq:cvalues}
c_1 = -2.60 \times 10^3 \mbox{, \ } 
c_2 = 5.02 \times 10^3 \mbox{, \ }
c_3 = -2.42 \times 10^3 
\end{equation} 
with units mol m$^{-4}$. The corresponding 
\[
r = \frac{d \rho}{dx}(0) = M_1 c_1 + M_2 c_2 + M_3 c_3 = -60.6 \mbox{\ kg m$^{-4}$}
\]
The main aim of this work is to show that these results are very similar to the diagonal Fick diffusivity model derived above. This will be done below. However, we also point out two things from the results above. The first is that the value of 
\[
c_1 + c_2 + c_3 = 0.2039
\]
is four orders of magnitude smaller than the individual values. In this isothermal situation, the sum above is proportional to the pressure gradient and this result shows that the bulk transport mechanism of Darcy flow is significantly more efficient than diffusion. This makes flow between channels in a serpentine flow field (in which adjacent channels can have a significant pressure drop between them) important to consider \cite{fick}. 

The second point to make is that over a representative GDL width $L$ of 100 microns, the results above predict an Oxygen concentration change of 
\[
c_1 L = -0.260
\]
which is less than a 2\% change in the channel concentration $C_1$. Even considering larger $L$ values (thicker GDL) or increasing $L$ with tortuousity effects \cite{tortpaper} does not make GDL transport losses significant. This is a reminder that current is not limited by dry channel and GDL transport but rather by processes in the catalyst layer, as is known in the literature (\cite{berg} for example). That this change is small over the GDL is predicted by the scaling arguments in the articles cited above. 

We proceed to computing $c_1$, $c_2$ and $r$ (the density derivative) using the diagonal Fick approximation of section~\ref{s:diagonal}. Proceeding as above, we can derive the system 

\begin{eqnarray*}
& & 
\left(
-\sigma 
\left[
\begin{array}{ccc}
C_1 & 0 & 0 \\
0 & C_2& 0  \\
0 & 0 & \rho
\end{array}
\right]
\mathcal{N} 
- \left[
\begin{array}{ccc}
D & 0 & 0 \\
0 & D & 0  \\
0 & 0 &0
\end{array}
\right]
+ \frac{D}{C}
\left[
\begin{array}{ccc}
C_1 & 0 & 0 \\
0 & C_2& 0  \\
0 & 0 & 0 
\end{array}
\right]
\mathcal{N}
\right)
\left[ \begin{array}{c} c_1 \\ c_2 \\ r \end{array} \right] \\
& & 
= 
\left[ \begin{array}{c} Q_1 \\ Q_2 \\ M_1Q_1 + M_2Q_2 + M_3Q_3 \end{array} \right]
\end{eqnarray*}
where
\[
\mathcal{N} = 
\left[
\begin{array}{ccc}
1-M_1/M_3 & 1-M_2/M_3 & 1/M_3 \\
1-M_1/M_3 & 1-M_2/M_3 & 1/M_3 \\
1-M_1/M_3 & 1-M_2/M_3 & 1/M_3 
\end{array}
\right]
\]
($\mathcal{N} [c_1, c_2, r]^T$ gives $c_1+c_2 + c_3$ in each component). The results with the same conditions as above are 
\[
c_1 = -2.49 \times 10^3 \mbox{, \ } 
c_2 = 4.81 \times 10^3 \mbox{, \ }
r = -58.0 
\]
As a reminder, these are the results using the diagonal Fick approximation. Comparing to the Maxwell Stefan values (\ref{eq:cvalues}) and the derived $r$, we see that the results differ by only a few percent. This was expected from the quality of the diagonal approximation in section \ref{s:diagonal}. 

\section{Summary}

The use of a diagonal Fick diffusivity is equivalent to the Maxwell Stefan diffusion model relative to a mass averaged velocity only when the molar masses of the species are identical and all  binary diffusivities are identical. However, in the case of a PEMFC cathode with (possibly compressed and humidified) ambient air as oxidant, it is shown that when oxygen and water vapour are used as the diffusing species, there is negligible difference in the results of the two models. Analytic insight into the result is gained with simple algebraic models. 

\section*{Acknowledgements}

The first author thanks NSERC for a graduate scholarship and the Automotive Fuel Cell Corporation (AFCC) and the MITACS Accelerate Internship programme for funding for this work. The second author acknowledges research funding support from an NSERC Canada grant. 

\bibliographystyle{plain}
\bibliography{bib}

\begin{thebibliography}{10}

\bibitem{Steam}
{\em International Steam Tables}.
\newblock Springer-Verlag, 2008.

\bibitem{CRC}
{\em CRC Handbook of Chemistry and Physics}.
\newblock CRS Press, 90th edition, 2009.

\bibitem{whittaker}
J.~Bear and Y.~Bachmat.
\newblock {\em Introduction to Modelling of Transport Phenomena in Porous
  Media}.
\newblock Kluwer Academic: Dordrecht, 1990.

\bibitem{berg}
P~Berg, K~Promislow, J~St~Pierre, J~Stumper, and B~Wetton.
\newblock {Water management in PEM fuel cells}.
\newblock {\em {JOURNAL OF THE ELECTROCHEMICAL SOCIETY}},
  {151}({3}):{A341--A353}, {MAR} {2004}.

\bibitem{review1}
Rod Borup, Jeremy Meyers, Bryan Pivovar, Yu~Seung Kim, Rangachary Mukundan,
  Nancy Garland, Deborah Myers, Mahlon Wilson, Fernando Garzon, David Wood,
  Piotr Zelenay, Karren More, Ken Stroh, Tom Zawodzinski, James Boncella,
  James~E. McGrath, Minoru Inaba, Kenji Miyatake, Michio Hori, Kenichiro Ota,
  Zempachi Ogumi, Seizo Miyata, Atsushi Nishikata, Zyun Siroma, Yoshiharu
  Uchimoto, Kazuaki Yasuda, Ken-ichi Kimijima, and Norio Iwashita.
\newblock {Scientific aspects of polymer electrolyte fuel cell durability and
  degradation}.
\newblock {\em {CHEMICAL REVIEWS}}, {107}({10}):{3904--3951}, {OCT} {2007}.

\bibitem{ptl}
O.S. Burheim, G.~Ellila, J.D. Fairweather, A.~Labouriau, S.~Kjelstrup, and J.G.
  Pharoah.
\newblock Ageing and thermal conductivity of porous transport layers used for
  \{PEM\} fuel cells.
\newblock {\em Journal of Power Sources}, 221(0):356 -- 365, 2013.

\bibitem{stack}
Paul Chang, Gwang-Soo Kim, Keith Promislow, and Brian Wetton.
\newblock {Reduced dimensional computational models of polymer electrolyte
  membrane fuel cell stacks}.
\newblock {\em {JOURNAL OF COMPUTATIONAL PHYSICS}}, {223}({2}):{797--821}, {MAY
  1} {2007}.

\bibitem{fick}
D.~H. Jeon, S.~Greenway, S.~Shimpalee, and J.~W. Van~Zee.
\newblock {The effect of serpentine flow-field designs on PEM fuel cell
  performance}.
\newblock {\em {INTERNATIONAL JOURNAL OF HYDROGEN ENERGY}},
  {33}({3}):{1052--1066}, {FEB} {2008}.

\bibitem{fms}
Michael~J. Martinez, Sirivatch Shimpalee, and J.~W. Van~Zee.
\newblock {Comparing predictions of PEM fuel cell behavior using Maxwell-Stefan
  and CFD approximation equations}.
\newblock {\em {COMPUTERS \& CHEMICAL ENGINEERING}}, {32}({12}):{2958--2965},
  {DEC 22} {2008}.

\bibitem{frank2}
Keith Promislow, Paul Chang, Herwig Haas, and Brian Wetton.
\newblock Two-phase unit cell model for slow transients in polymer electrolyte
  membrane fuel cells.
\newblock {\em Journal of The Electrochemical Society}, 155(7):A494--A504,
  2008.

\bibitem{keith}
Keith Promislow, John Stockie, and Brian Wetton.
\newblock A sharp interface reduction for multiphase transport in a porous fuel
  cell electrode.
\newblock {\em Proceedings of the Royal Society A: Mathematical, Physical and
  Engineering Science}, 462(2067):789--816, 2006.

\bibitem{stockiepaper}
J.~Stockie, K.~Promislow, and B.~Wetton.
\newblock {A Finite Volume Method for Multicomponent Gas Transport in a Porous
  Fuel Cell Electrode}.
\newblock {\em {International Journal for Numerical Methods in Fluids}},
  {462}:{789--186}, {2003}.

\bibitem{TK}
M.~Taylor and R.~Krishna.
\newblock {\em Multicomponent Mass Transfer}.
\newblock John Wiley \& Sons, 1993.

\bibitem{tortpaper}
Xiao-Dong Wang, Jin-Liang Xu, and Duu-Jong Lee.
\newblock Parameter sensitivity examination for a complete three-dimensional,
  two-phase, non-isothermal model of polymer electrolyte membrane fuel cell.
\newblock {\em International Journal of Hydrogen Energy}, 37(20):15766 --
  15777, 2012.
\newblock The 2011 Asian Bio-Hydrogen and Biorefinery Symposium (2011ABBS).

\bibitem{wilkinson}
D.P. Wilkinson, J.~Zhang, R.~Hui, J.~Fergus, and X~Li.
\newblock {\em Proton Exchange Membrane Fuel Cells: Materials Properties and
  Performance}.
\newblock Green Chemistry and Chemical Engineering. CRC Press, 2009.

\bibitem{ms}
Woo-Joo Yang, Hong-Yang Wang, and Young-Bae Kim.
\newblock Effects of the humidity and the land ratio of channel and rib in the
  serpentine three-dimensional pemfc model.
\newblock {\em International Journal of Energy Research}, 37(11):1339--1348,
  2013.

\end{thebibliography}
\end{document}